\documentclass[aps,prd,twocolumn,showpacs,floatfix]{revtex4-1}
\usepackage{epsfig}
\usepackage{multirow}
\usepackage{graphicx}
\usepackage{subfigure}

\begin{document}

\title{Higgs boson hadronic branching ratios at the ILC}

\author{Yambazi Banda}\affiliation{University of Oxford,\\ Denys Wilkinson Building, Keble Road,\\ Oxford OX1 3RH, UK}
\author{Tom\'{a}\v{s} La\v{s}tovi\v{c}ka}\affiliation{University of Oxford,\\ Denys Wilkinson Building, Keble Road,\\ Oxford OX1 3RH, UK}
\author{Andrei Nomerotski}\affiliation{University of Oxford,\\ Denys Wilkinson Building, Keble Road,\\ Oxford OX1 3RH, UK}

\begin{abstract}

We present a study of the Higgs boson decay branching ratios to $b\bar{b}$, $c\bar{c}$ and gluons, one of the cornerstones of the physics %%@
program at the International Linear Collider (ILC). A standard model Higgs boson of 120\,GeV mass, produced in the Higgs-strahlung process at %%@
\mbox{$\sqrt{s} = 250$\,GeV} was investigated using the full detector simulation and reconstruction procedures. The analysis was performed in %%@
the framework of the Silicon Detector (SiD) concept with full account of inclusive standard model backgrounds. The selected decay modes %%@
contained two heavy flavour jets in the final state and required excellent flavour tagging through precise reconstruction of interaction and %%@
decay vertices in the detector. A new signal discrimination technique using correlations of neural network outputs was used to determine the %%@
branching ratios and estimate their uncertainties, 4.8\%, 8.4\% and 12.2\% for $b\bar{b}$, $c\bar{c}$ and gluons respectively.

\end{abstract}


\maketitle

% -----------------------------------------------------------------------------------------------

\section{Introduction}

The measurement of the Higgs absolute branching ratios in all possible decay modes is one of the most important parts of the ILC program, %%@
giving a precision test of the standard model (SM) prediction that the Higgs boson couples to each particle in proportion to its %%@
mass~\cite{Higgs}. While the Large Hadron Collider (LHC) is a likely discovery machine for the Higgs boson it will be almost impossible to %%@
determine the total Higgs cross section there, making it difficult to make absolute measurements of partial decay widths. On the other hand, %%@
precise and model independent measurements of absolute branching ratios can be performed at the ILC \cite{lhcilc}.

In this paper the Higgs decay modes to charm quarks, bottom quarks and gluons which result in two and four-jet final states are considered.
Heavy quarks, $b$ and $c$, couple directly to the Higgs while the Higgs decay to gluons in the standard model is mediated by heavy quark %%@
loops. The branching ratio to gluons is indirectly related to $t\bar{t}H$ Yukawa coupling~\cite{yukawa} and would probe the existence of new %%@
strongly interacting particles that couple to the Higgs and are too heavy to be produced directly. The precise measurements of the branching %%@
ratios will help to discriminate between different `Beyond the SM' (BSM) scenarios~\cite{mssm2,mssm1} where the Higgs couplings differ from SM %%@
Higgs couplings.

The production cross sections, branching ratios and decay widths of the Higgs are determined by the strength of the Yukawa couplings to %%@
fermions and gauge bosons, whose scale is set by the masses of these particles. 
For a Higgs boson mass of 120 GeV, the expected branching ratios to bottom quarks, charm quarks and gluons are  67.9\%, 3.1\% and 7.1\%, %%@
respectively.

While the Higgs branching ratios in the $e^+e^-$ collider environment has been studied before \cite{ZH}, up to date it was done in a %%@
parametric way from the detector point of view, without exploiting full Monte Carlo (MC) simulations nor realistic reconstruction algorithms. %%@
It should be noted that this measurement is a complex analysis which heavily relies on the bottom and charm quark tagging. Possible detector %%@
resolution and acceptance effects together with non-ideal efficiency and purity of the reconstruction algorithms could play a critical role in %%@
the ultimate sensitivity of the experiment and hence in its physics reach. This study for the first time addresses these issues in the %%@
framework of the Silicon Detector (SiD) concept as one of the benchmarking analyses prepared for the SiD Letter of Intent \cite{LoI}. A new %%@
technique employing two neural networks was used to overcome one of the main difficulties of the analysis: separation of signal from the %%@
background of Higgs decays to other particles. 

The paper is organized as follows: the Higgs boson production at the ILC and the SiD detector concept are introduced in Section~\ref{ilc_sid}; %%@
Sections~\ref{analysis} and~\ref{res} describe the details of the analysis and its results, which are summarized in Section~\ref{summary}.

\section{Higgs production at the ILC}\label{ilc_sid}

\subsection{The ILC}

The ILC is a next generation electron-positron accelerator designed to collide particles at the centre-of-mass energy up to 500\,GeV, 
upgradeable to 1\,TeV, with peak luminosity of 2$\times$10$^{34}$\,cm$^{-2}$s$^{-1}$. The dominant production mechanisms for the SM Higgs at %%@
$e^+e^-$ colliders are Higgs-strahlung~\cite{zhggstr,zhggstr2}, $WW$ and $ZZ$ fusion processes~\cite{fuse}

\begin{equation}
\begin{array}{lll}
e^+e^- \to ZH \to f\bar{f}H \\
e^+e^- \to \nu\bar{\nu}W^{\ast}W^{\ast} \to \nu\bar{\nu}H \\
e^+e^- \to e^+e^-Z^{\ast}Z^{\ast} \to e^+e^-H.
\end{array}
\end{equation}

The Higgs-strahlung and $WW$ fusion processes are both capable of generating $\nu\bar{\nu}H$ states.

\subsection{The SiD Detector Concept}

\begin{figure*}
\centerline{\includegraphics[width=\linewidth]{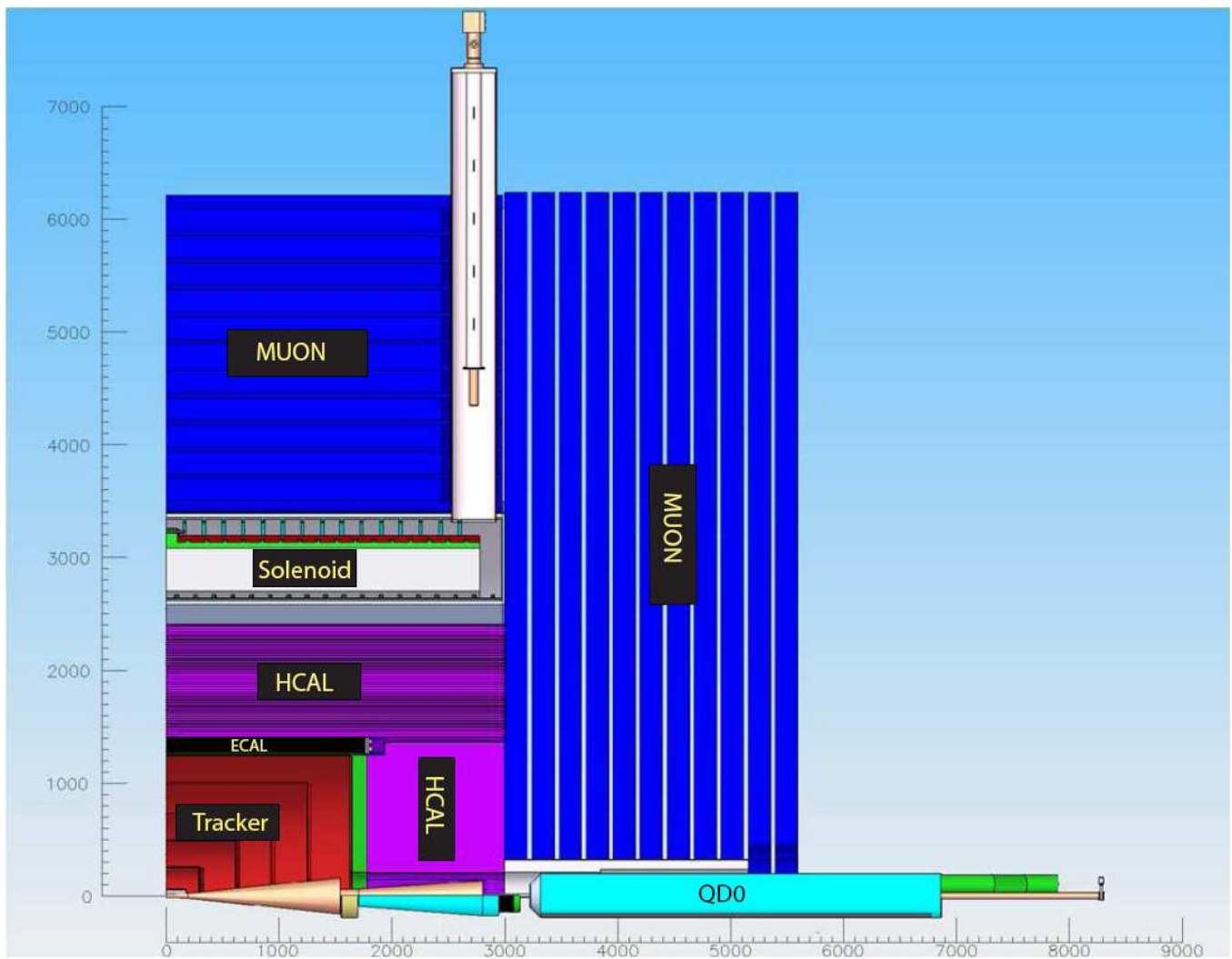}}
\caption{A plane view of a quadrant of SiD. Dimensions are in mm.}
\label{fig:SiD}
\end{figure*}

The sensitivity to Higgs decay branching ratios measurement was studied in the framework of the Silicon Detector (SiD) concept~\cite{LoI} %%@
using full detector simulation and event reconstruction. SiD is designed for precision measurements of a wide range of possible new phenomena %%@
at the ILC. It is based on a silicon pixel vertex detector, silicon tracking, silicon-tungsten electromagnetic calorimetry, and highly %%@
segmented hadronic calorimetry. The Particle Flow Algorithm (PFA) approach~\cite{LoI} is an important strategy driving the basic philosophy %%@
and layout of the detector. SiD also incorporates a 5\,T solenoid, iron flux return and a muon identification system. A schematic view of SiD %%@
quadrant is shown in Figure~\ref{fig:SiD}.

The vertex detector comprises a central barrel with five silicon pixel layers and four silicon pixel disks for the forward and backward %%@
regions providing the impact parameter resolution of 
$5  \oplus 10/(p \sin^{3/2}\theta$)\ [$\mu$m] and good hermeticity for \mbox{$\cos \theta$ $\le$ 0.984} where \mbox{$\theta$} is the polar %%@
angle.

SiD has adopted silicon strip technology for its tracker, arranged in five cylinders and four endcaps necessary for precision tracking and %%@
momentum measurement. Given an outer cylinder radius of 1.25\,m, the charged track momentum resolution is expected to be better than %%@
$\sigma(1/p_T) = 5\, \times 10^{-5}$ (GeV/c)$^{-1}$ for high momentum tracks.
 
Calorimetry in SiD is optimized for jet energy measurements, and is based on the Particle Flow strategy, in which the momenta of charged %%@
particles are measured in the tracker; the momenta of neutrals are measured in the calorimeters, and then the charged and neutral components %%@
are combined.
The SiD calorimetry begins with a dense, highly pixelated silicon-tungsten electromagnetic calorimeter with  energy resolution of %%@
17\%/$\sqrt{E}$. The Hadronic Calorimeter (HCAL) is composed of 4.5$\lambda$ HCAL depth of stainless steel, divided into 40 layers of steel %%@
and detector.
A good jet energy resolution, about 4\%, allows
the invariant masses of $W$'s, $Z$'s and top quarks to be reconstructed with resolution nearing the natural widths of these particles.

\section{Analysis}\label{analysis}

\subsection{Analysis Framework}

Both the signal and background events were produced at the centre-of-mass energy $\sqrt{s}=250$~GeV, total integrated luminosity of
250~fb$^{-1}$ and the Higgs mass of 120 GeV. The choice of energy in this analysis maximizes the cross-section value for
Higgs-strahlung. Only $HZ$ process was considered as it accounts for about 95\% of the total $Hf\bar{f}$ cross section for this energy and %%@
Higgs mass choices. SM events (mainly $WW$, $ZZ$ and $qq$ pairs) and Higgs decays to fermions other than the signal were considered as %%@
backgrounds.  All 0, 2 and 4 fermion final states were generated using the WHIZARD Monte Carlo Event Generator~\cite{whiz} taking into account %%@
beamstrahlung and initial state radiation (ISR) photons~\cite{LoI}. PYTHIA~\cite{pyth} was used for the final state QED and QCD parton %%@
showering, fragmentation and decay to provide final-state observable particles. For this study, event samples were created conforming to the %%@
expected ILC baseline parameters of $\pm$80\% electron and $\mp$30\% positron polarization. About 7M events were processed through the full %%@
detector simulation, with individual events weighted to reflect the statistical sampling. 

The detector response to generated events was simulated using the Geant4 toolkit~\cite{geant,geant2}, which provided the necessary classes
to describe the geometry of the detector, the transport and interaction of particles with materials and fields. A thin layer of 
Linear Collider specific code, SLIC~\cite{slic}, provided access to the Monte Carlo events, the detector geometry and 
the output of the detector hits. The detector parameters could be varied without having to rebuild the simulation executable binaries
since the geometries were fully described at runtime. The output was in the standard LCIO format~\cite{lcio} so that detector concepts using
other simulation packages could be studied and data generated using this system can be analyzed in other analysis frameworks. 

The identification of jets is an important part of this analysis. The fragmentation products of the hadronic systems were forced either to two %%@
or four jets, depending on the final state, using
the DURHAM algorithm~\cite{durham}. To provide the most probable kinematic configuration of the event topology, a kinematic fitter, Marlin %%@
Kinfit~\cite{kinfit}, with four-momenta and mass
constraints was used. The fitter uses the method of Lagrange multipliers to determine the most probable value for the jet four-momentum.

In order to identify primary, secondary and tertiary vertices the topological vertex finder ZVTOP was utilized. The algorithm
is part of a vertexing package developed by the LCFI collaboration~\cite{lcfi}. It classifies events
on the number of found vertices and combines eight optimized variables for each type of event in a neural network, which is then
separately trained on samples of $b$, $c$ and light quarks. The best discriminating variables were the corrected vertex mass, secondary
vertex probability, impact parameter significance and the number of vertices in the event.

\subsection{Event Selection}

The analysis signature is dependent on the $Z$ boson decay products (charged leptons, hadrons or neutrinos). The channels studied in this
analysis were the neutrino mode ($Z$ decaying to neutrinos) and the hadronic mode ($Z$ decaying to hadrons), referred below as the signal.

\begin{figure*}[htpb]
\subfigure[]{\includegraphics[width=0.48\linewidth]{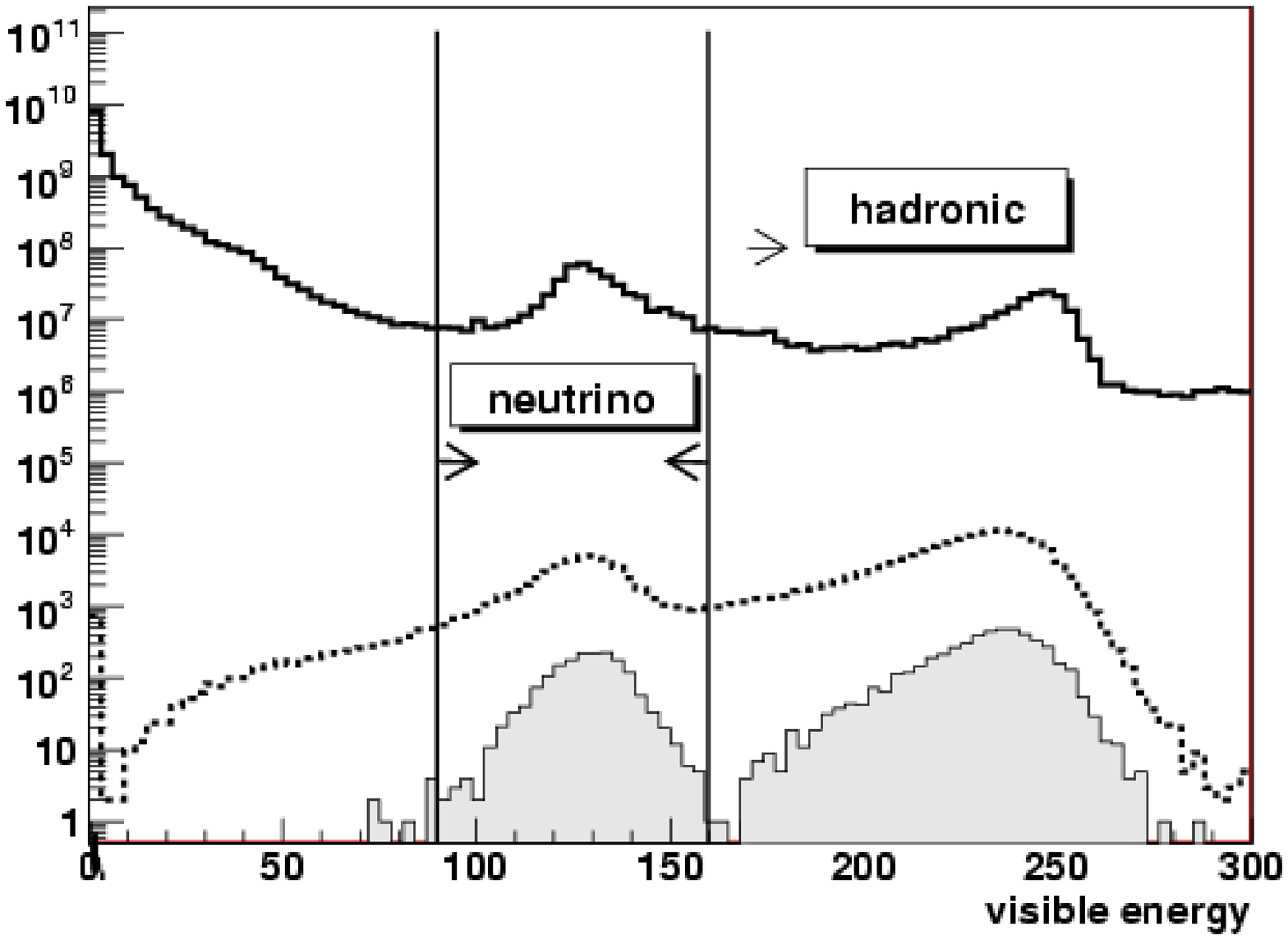}}
\subfigure[]{\includegraphics[width=0.48\linewidth]{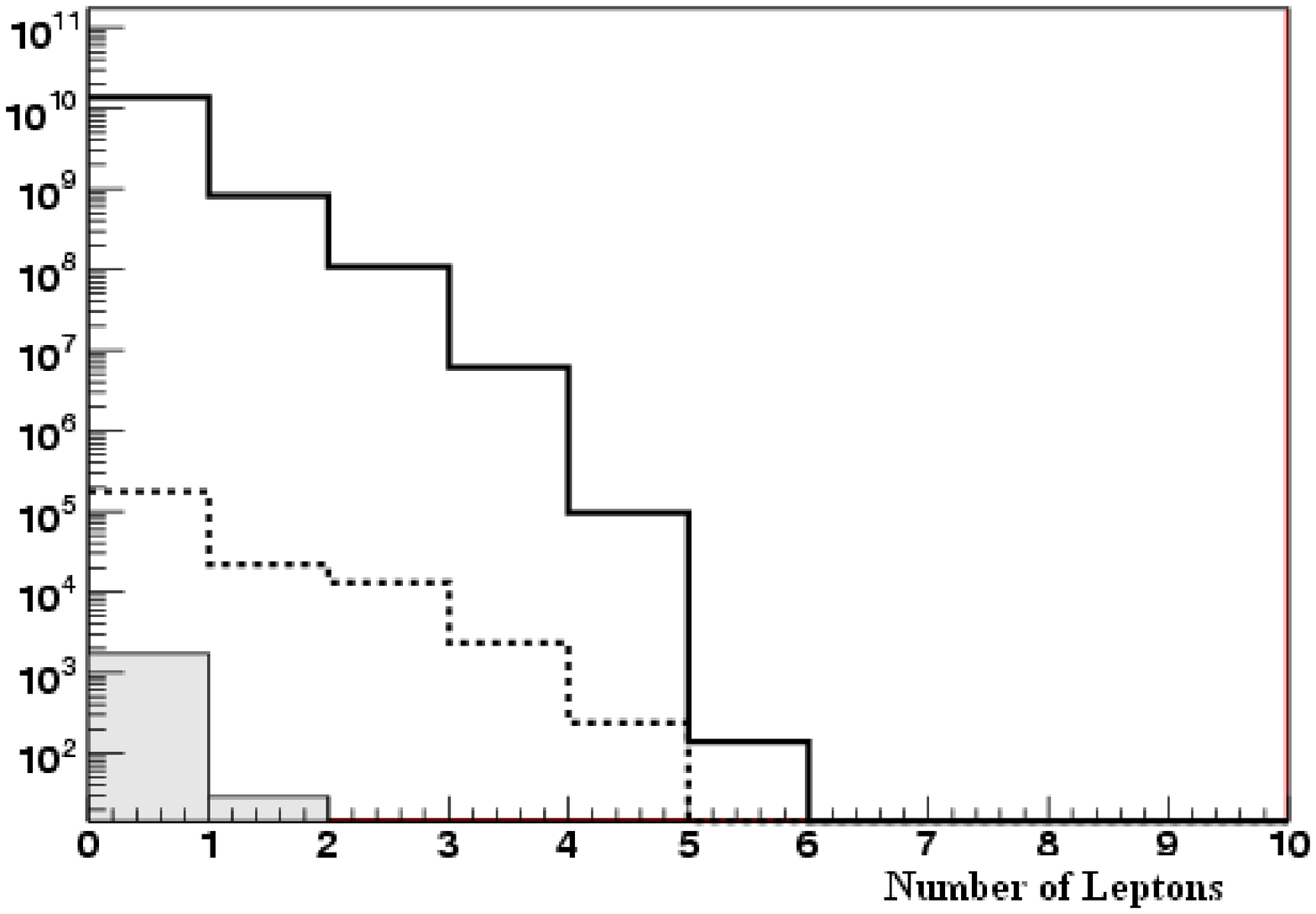}}
\caption{\label{fig:Enlep}
Visible energy (a) and the number of leptons per event (b) used for channel classification. Solid curves are SM background, dashed curves are %%@
inclusive Higgs sample and filled histograms are the signals. All histograms are normalized to 250\,fb$^{-1}$.}
\end{figure*}

The events were classified into the two channels using the number of leptons and visible energy in the event. The leptons were defined as %%@
reconstructed electrons or muons with momenta larger than 15\,GeV. The visible energy was defined as the sum of energies of all reconstructed %%@
particles~\cite{LoI} in the event. Figure~\ref{fig:Enlep} shows the distributions of the visible energy and the number of leptons for the %%@
signal, Higgs background and SM background before any selections, normalized to 250\,fb$^{-1}$.

In the neutrino mode, events not having leptons and with visible energy in the \,90 to 160\,GeV interval were selected.
In this channel all reconstructed particles were clustered into two jets which, for the signal, were assumed to come from the Higgs boson %%@
recoiling against two neutrinos from the $Z$ boson decay. The most important backgrounds in this channel were two-fermion events, $ZZ$ pairs %%@
decaying to neutrinos and hadrons, and $WW$ pairs where one $W$ decays hadronically and the other $W$ decays into a neutrino and a lepton, %%@
which escapes undetected. 

\begin{figure*}[t]
\subfigure[]{\includegraphics[width=0.48\linewidth]{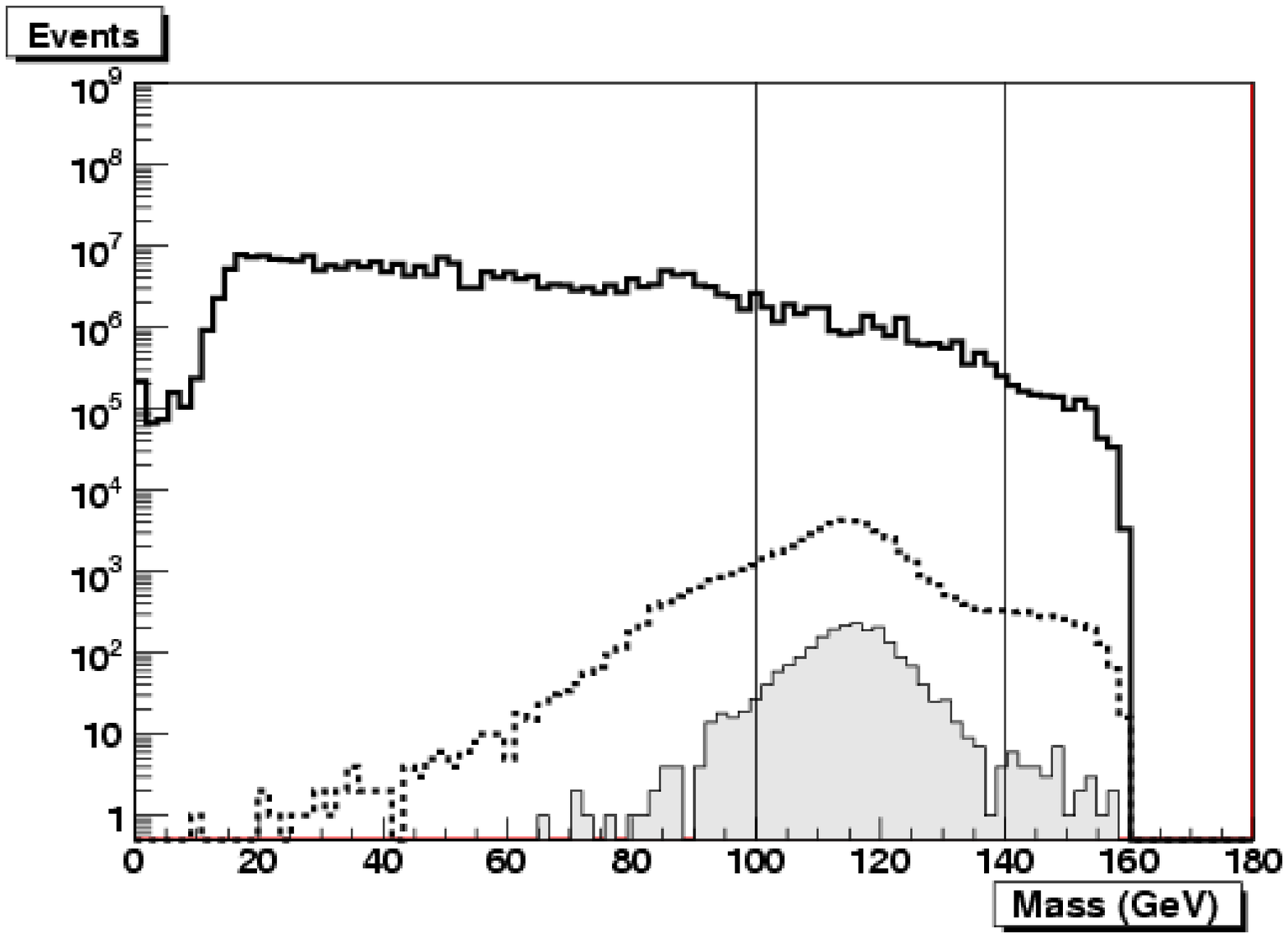}}
\subfigure[]{\includegraphics[width=0.48\linewidth]{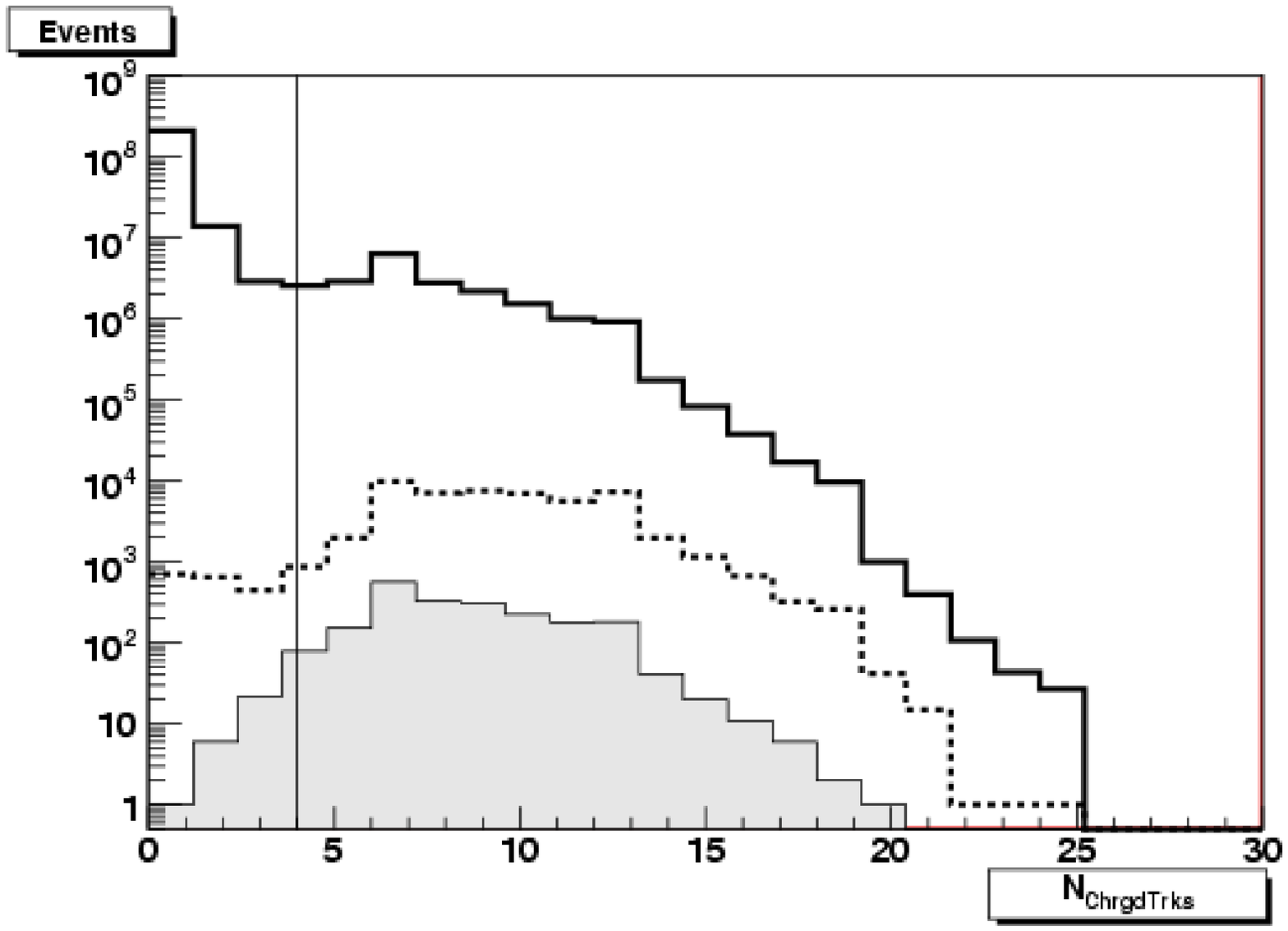}} \\
\subfigure[]{\includegraphics[width=0.48\linewidth]{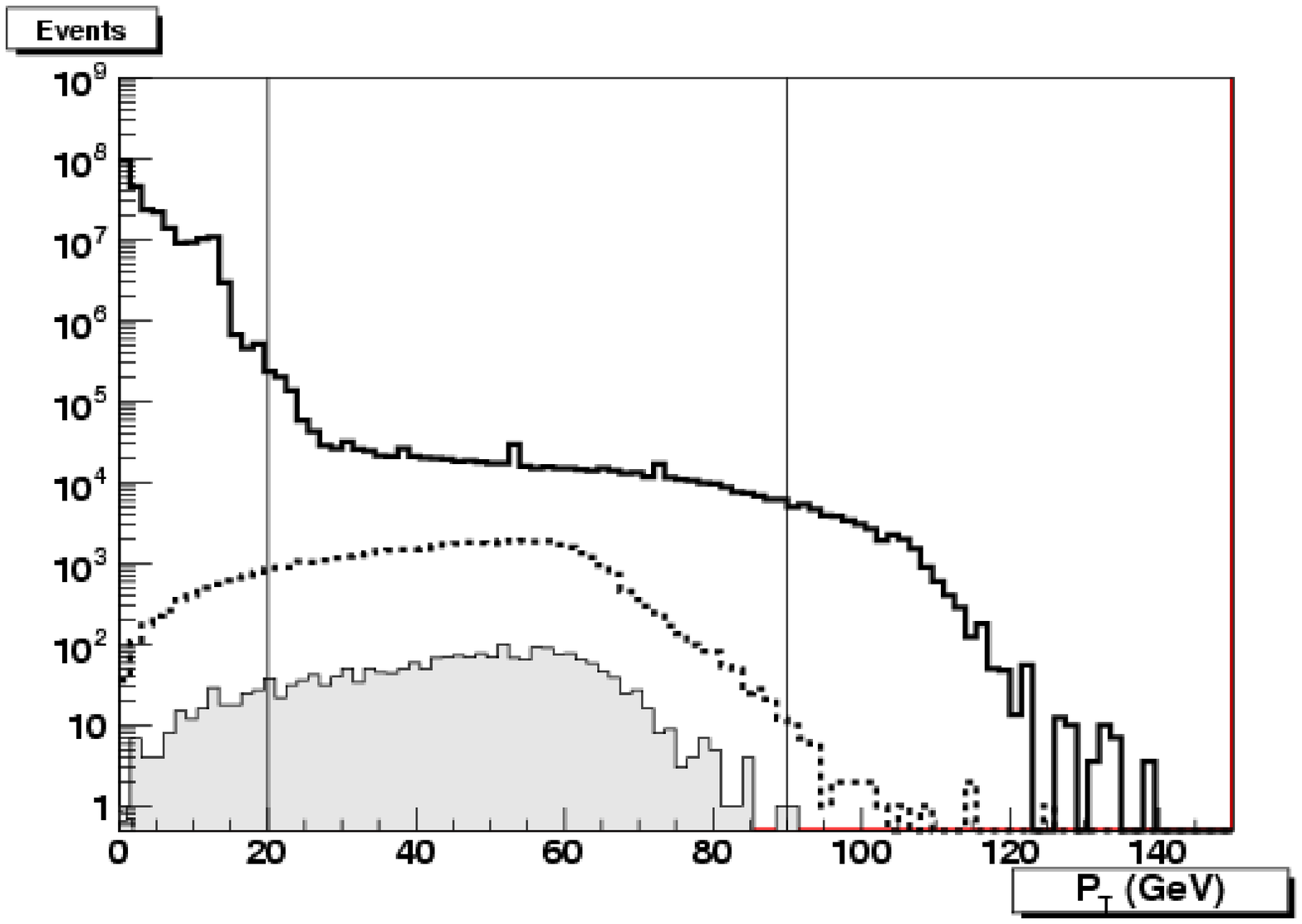}}
\subfigure[]{\includegraphics[width=0.48\linewidth]{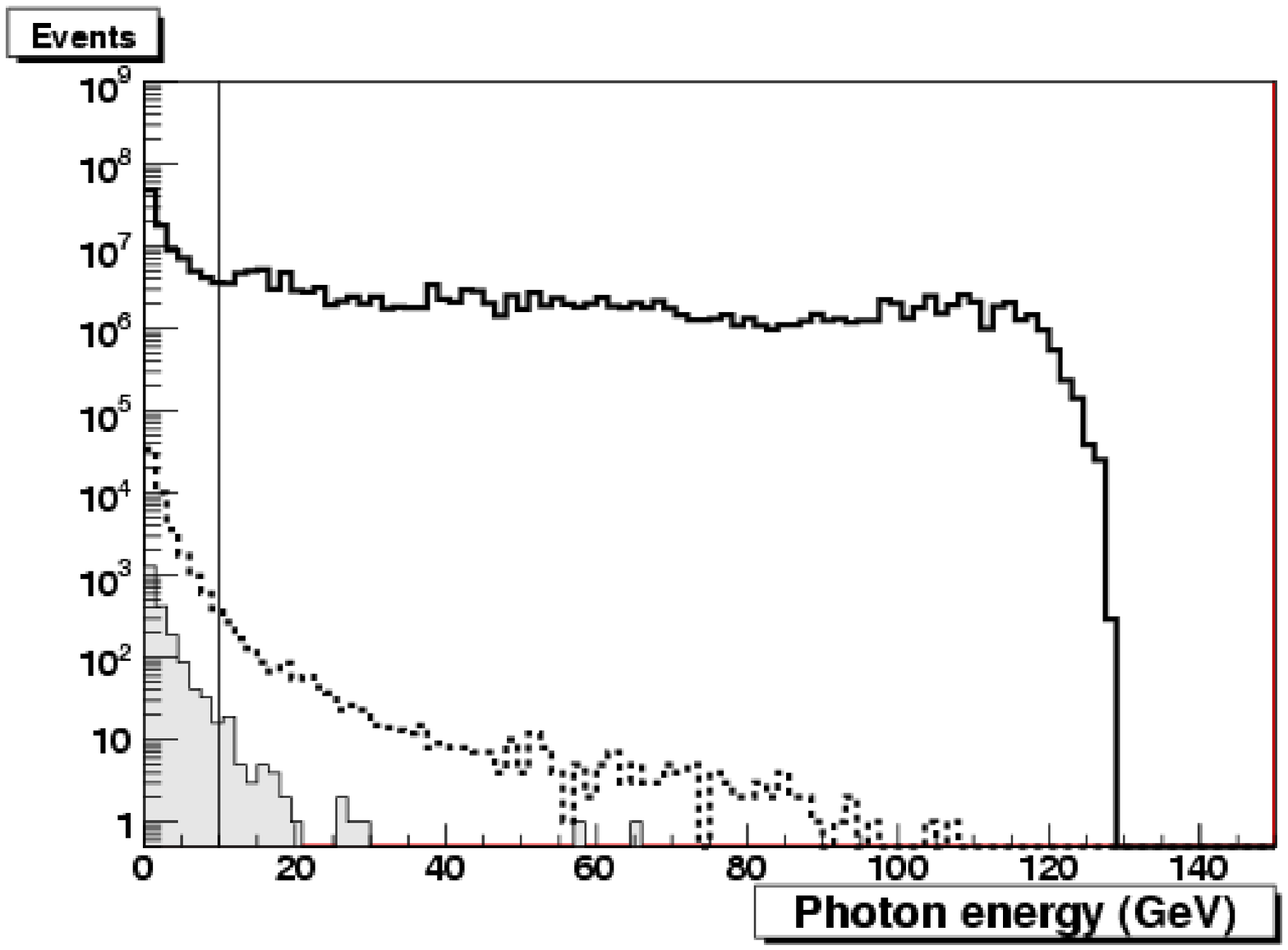}}
\caption{\label{cutfigs}
Distributions of pre-selection kinematic variables in the neutrino channel: (a) di-jet invariant mass, (b) number of charged tracks in the %%@
event, (c) jet p$_T$, (d) highest reconstructed photon energy in the event. Solid curves are the SM background, dashed curves are the Higgs
background and filled histograms represent the signal. All histograms are normalized to 250\,fb$^{-1}$ and are shown before any selections.}
\end{figure*}

A cut based selection was performed to further reduce the backgrounds in this channel using kinematic variables described below. The %%@
transverse momentum of each jet, $p_T$, was calculated to reject SM background events which are softer compared to signal events. The number %%@
of tracks per jet, $n_{tracks}$, was used to reject purely leptonic events which were not part of the signal. The discrimination between the %%@
signal and backgrounds with a different number of jets was achieved by the $y_{min}$ parameter, which corresponds to the minimum $y$-parameter %%@
for the two jet hypothesis in the DURHAM algorithm. The variable $-\log(y_{min}$) was used to exclude fully hadronic $WW$ (and $ZZ$) events, %%@
which were four-jet events mis-identified as two-jet events. Thrust~\cite{thrs} values were calculated since it is expected that signal events %%@
are less boosted and are more spherical than the background events. Values of $\mid$$\cos(\theta_{thrust})$$\mid$ were also used because %%@
signal events are expected to be produced more centrally in the detector while majority of the background processes
have a strongly forward peaking angular distribution. A large fraction of background events have back-to-back jets whereas signal events were %%@
confined to a range of angular values due to kinematic constraints coming from $ZH$ production. The di-jet invariant mass was used to reject %%@
low mass hadronic systems from $WW$, $ZZ$ and two-photon events. The signal events rarely have hard photons and this helps reject both highly %%@
energetic initial state radiation (ISR) photons and  hard photons from di-jet events which occur for the background. No photon isolation was %%@
required in this particular case. Figure~\ref{cutfigs} shows some of the distributions for the above variables used in the neutrino channel %%@
before any selections.

The hadronic mode was selected as events with no leptons and the visible energy above 170\,GeV. In this channel, events were forced to have %%@
four reconstructed jets. For the signal, two of the jets were required to have their di-jet invariant mass consistent with the Higgs boson and %%@
the other two having the mass consistent with the $Z$ boson mass. The main backgrounds for this channel were $WW$ and $ZZ$ pairs where the all %%@
bosons decay to hadrons.
For the pre-selection, we used kinematic selections defined similar to the ones in the neutrino channel. The full list of selections in the %%@
neutrino and hadronic channels is given in Tables~\ref{tab:HZ1} and ~\ref{tab:HZ2}.  Table~\ref{NUMS} instead shows the number of events %%@
before and after pre-selection cuts for $c\bar{c}$, $b\bar{b}$ and $gg$ modes in the neutrino and hadronic channels before and after %%@
pre-selection cuts. The number of events for the SM background is indicated in the same table.  

\begin{center}
\begin{table*}
\begin{tabular}{|lllll|} \hline 
& && selection & value  \\ \hline
(1) &  20 & $<$ & p$_T$ of jet & $<$ 90 GeV \\
(2) &  4 & $<$ & number of charged tracks per jet & \\
(3) & && $-\log(y_{min}$) & $<$ 0.8 \\
(4) & && thrust & $< 0.95$ \\
(5) & && $ \cos(\theta_{thrust}) $&$< 0.98$ \\
(6) & $100^{\circ}$ &$<$ & angle between jets &$< 170^{\circ}$ \\
(7) & 100 GeV &$<$ & di-jet invariant mass &$<$ 140 GeV \\
(8) & && Highest reconstructed photon energy & $<$ 10 GeV \\
\hline
\end{tabular}
\caption{\label{tab:HZ1}
Selections for the neutrino channel.}
\end{table*}
\end{center}

\begin{center}
\begin{table*}
\begin{tabular}{|lllll|} \hline 
& && selection & value  \\ \hline
(1) & 4 & $<$ & number of charged tracks per jet &\\
(2) & && $-\log(y_{min}$) & $<$ 2.7 \\
(3) & && thrust & $< 0.95$ \\
(4) & && $ \cos(\theta_{thrust})$ & $< 0.96$ \\
(5) & $105^{\circ}$ &$<$ & angle between jet 1 and 3 &$< 165^{\circ}$ \\
(6) & $70^{\circ}$ &$<$ & angle between jet 2 and 4 &$< 160^{\circ}$ \\
(7) & 110 GeV &$<$ & invariant mass of Higgs candidate after fit &$<$ 140 GeV \\
(8) & 80 GeV &$<$ & invariant mass of $Z$ candidate after fit &$<$ 110 GeV \\
(9) & && Highest reconstructed photon energy & $<$ 10 GeV \\
\hline
\end{tabular}
\caption{\label{tab:HZ2}
Selections for the hadronic channel. The jets are ranked in the decreasing energy order.}
\end{table*}
\end{center}

\begin{figure*}[t]
\subfigure[]{\includegraphics[width=0.48\linewidth]{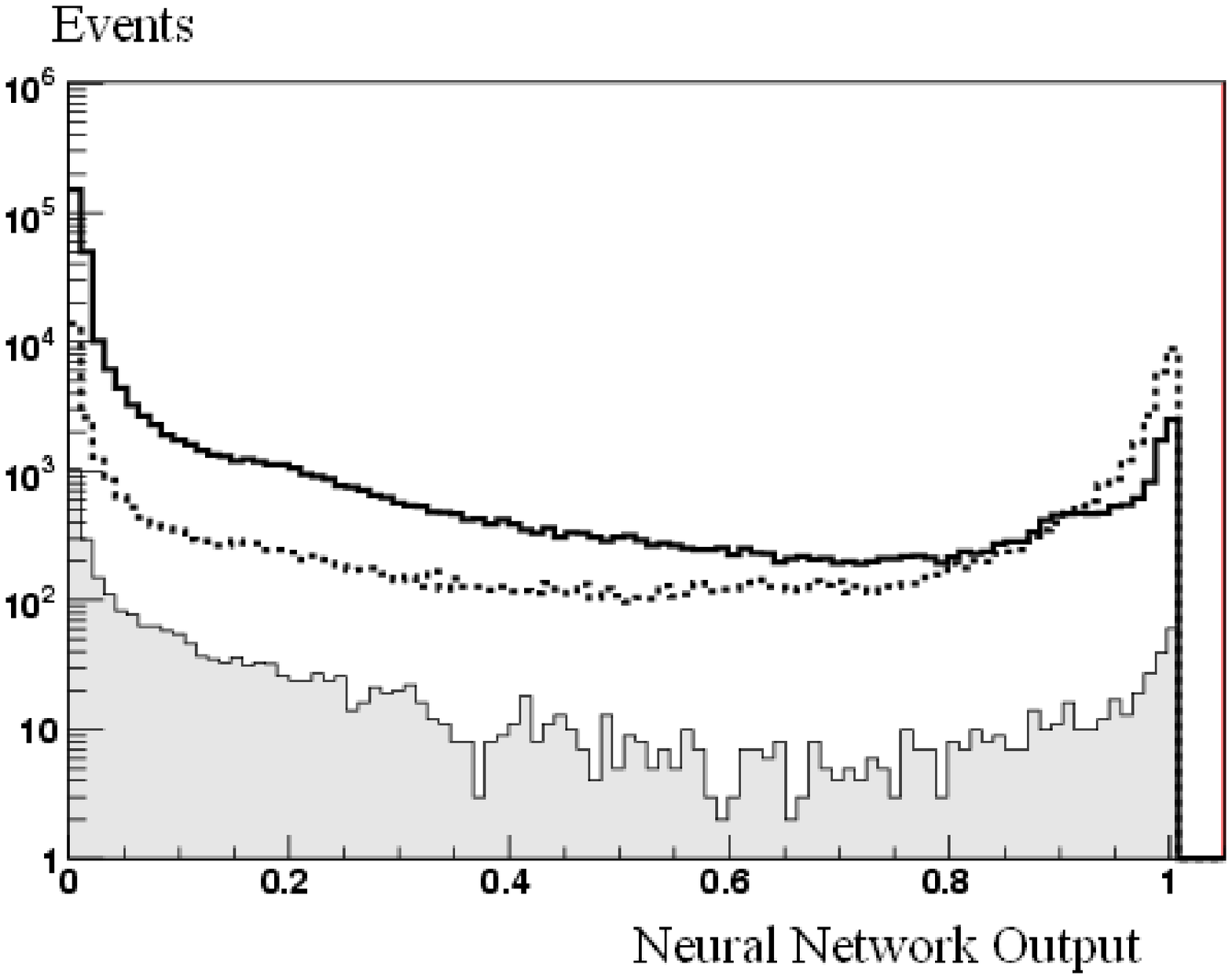}}
\subfigure[]{\includegraphics[width=0.48\linewidth]{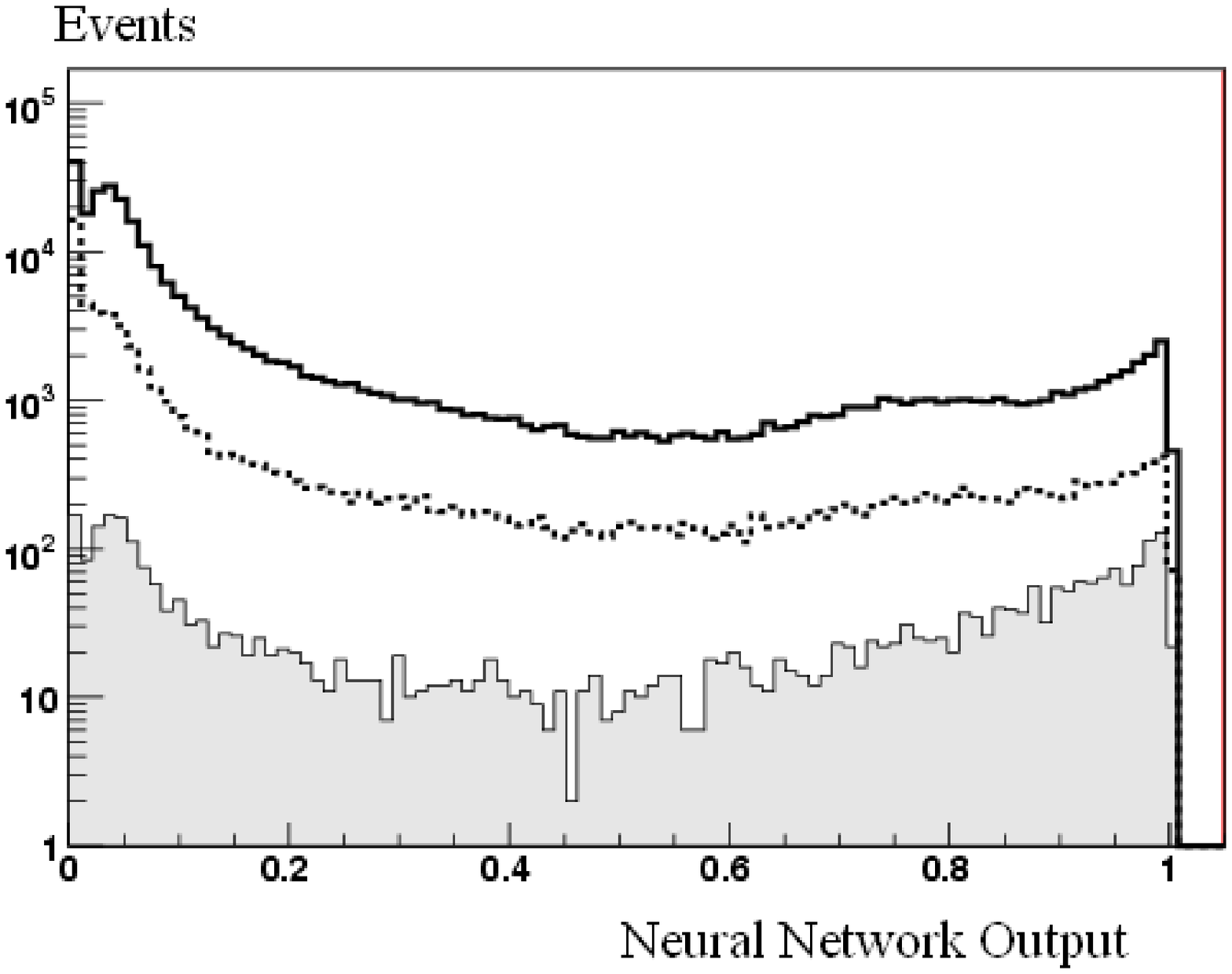}} \\
\subfigure[]{\includegraphics[width=0.48\linewidth]{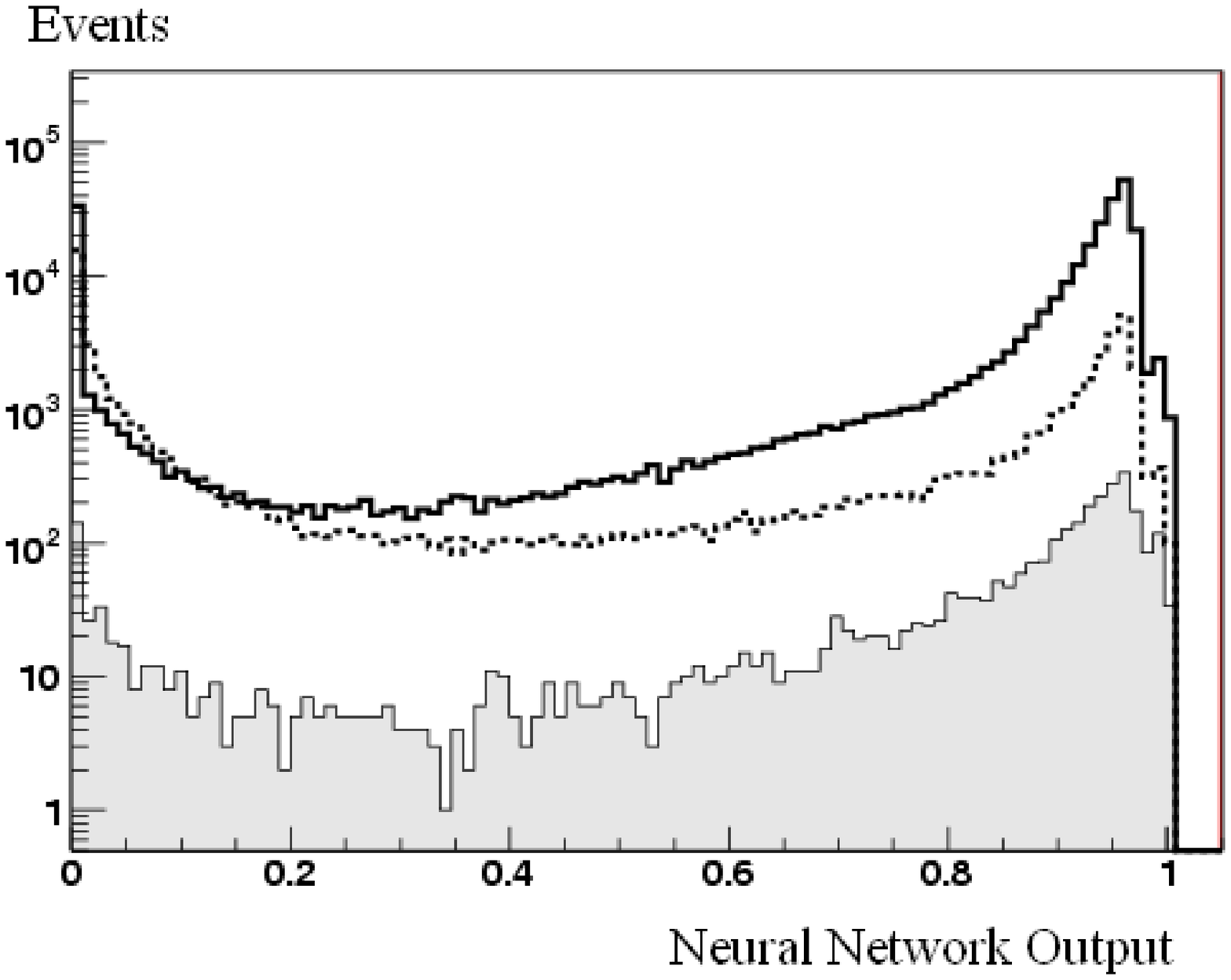}}
\caption{\label{fig:ccNN}
LCFI flavour tag distributions for the leading jet in the hadronic channel for the $c\bar{c}$ decay mode: (a) $b$-tagging variable, (b) %%@
$c$-tagging variable and (c) $c$-tagging variable for the case of $b$-background only. Solid curves are SM background, dashed curves are Higgs %%@
background sample and filled histograms are the signal. All histograms are normalized to 250\,fb$^{-1}$ and are show after preselections.}
\end{figure*}

\begin{center}
\begin{table}
\begin{tabular}{| c | c | c | c | c | c | }
\hline
\multicolumn{2}{|c|}{~}	
& SM & $H \rightarrow c\bar{c}$ & $H \rightarrow b\bar{b}$ & $H \rightarrow gg$ \\
\hline
\multirow{2}{*}{Neutrino Channel}
& Before & 45936973 & 637 & 11580 & 986 \\
& After  & 109057 & 506 & 6707 & 759 \\
\hline
\multirow{2}{*}{Hadronic Channel}
& Before & 39398366 & 1837 & 30985 & 2965 \\
& After  & 967312 & 947 & 15805 & 1611 \\
\hline	
\end{tabular}
\caption{Number of events before and after pre-selections in the neutrino and hadronic modes.}
\label{NUMS}
\end{table}
\end{center}
As mentioned before, the LCFI vertexing package was used for vertex finding and identification of the flavour of hadrons in jets. The flavour %%@
identification is essential to differentiate $b$ jets from $c$ jets and jets coming from light ($u, d, s$) quark hadronization. In analyses %%@
like the present one where multi-$b$ states are to be separated from significant backgrounds, a high $b$-tagging efficiency is required to %%@
discriminate against light and $c$-quarks. Furthermore, the measurement of the $H \rightarrow c\bar{c}$ branching ratio requires efficient %%@
$c$-tagging with high rejection of $b$-quarks. The package produces three possible flavour tags, `$b$-tag', `$c$-tag' and `$c$-tag on %%@
$b$-background only', for each jet in the event. Figure~\ref{fig:ccNN} shows the distribution of these variables for the leading jet in the %%@
hadron channel for the $c\bar{c}$ decay mode after preselections.

\begin{figure*}
\subfigure[]{\includegraphics[width=0.48\linewidth]{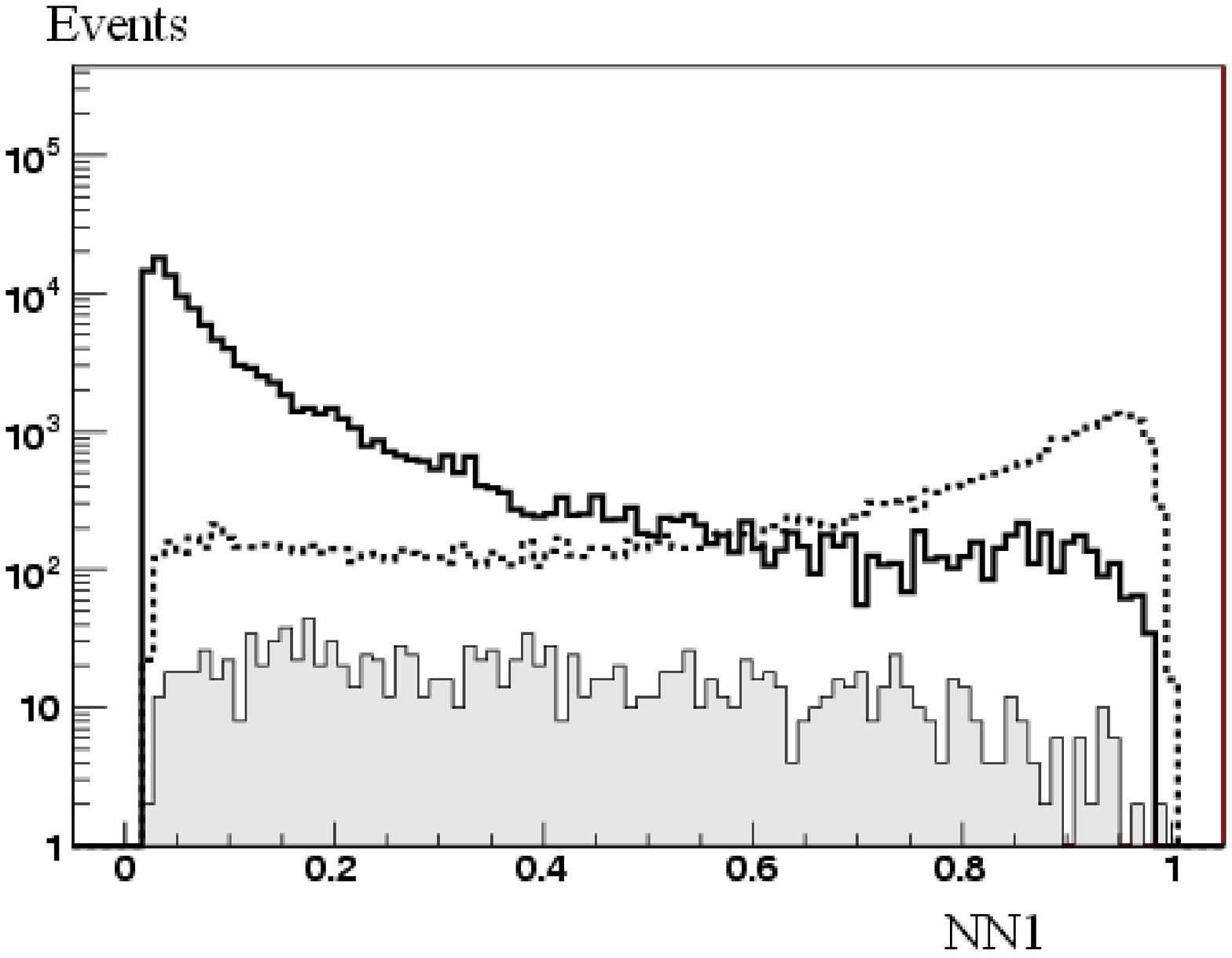}}
\subfigure[]{\includegraphics[width=0.48\linewidth]{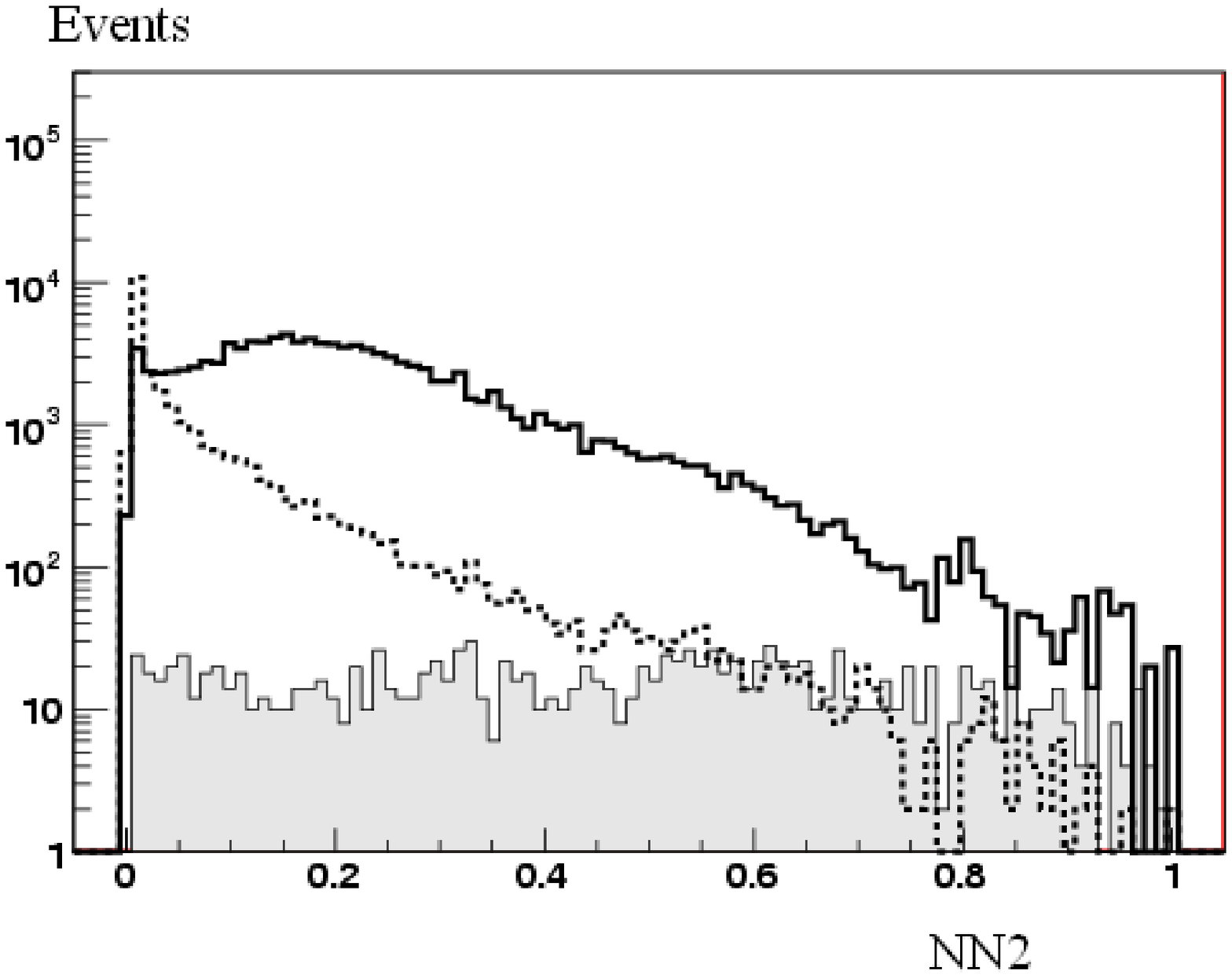}}
\caption{\label{fig:nns}
Neural network outputs in the $c\bar{c}$ decay mode for the neutrino channel: (a) First NN and (b) second NN. Solid curves are SM background, %%@
dashed curves are Higgs background sample and filled histograms are the signal. All histograms are normalized to 250\,fb$^{-1}$.}
\end{figure*}

The remaining events were categorized using the neural networks implemented in FANN~\cite{fann}. The neural network (NN) used optimized sets %%@
of 18 input variables for the neutrino channel and of 19 input variables for the hadronic channel. The neural net was constructed out of three %%@
hidden layers with 28 neurons and had one output neuron. 
For $c\bar{c}$ and $b\bar{b}$ signals, the first NN was trained to distinguish the SM background from the inclusive Higgs sample and to %%@
produce the NN$_{SM-Higgs}$ output. In the $gg$ case, the first NN was trained to distinguish the signal sample from the SM background and to %%@
produce the NN$_{Sig-SM}$ output. The second NN was, in all cases, trained to distinguish the signal from the Higgs inclusive background %%@
sample and to produce the NN$_{Higgs-signal}$ output. The training was done separately for $c\bar{c}$, $b\bar{b}$ and for $gg$ using %%@
independent samples. Figure~\ref{fig:nns} shows an example of distributions of the first and second NNs in the $c\bar{c}$ decay mode for the %%@
neutrino channel. For the $c\bar{c}$/$b\bar{b}$/$gg$ scenarios the signal was defined as $H\rightarrow$ $c\bar{c}$/$b\bar{b}$/$gg$ events only %%@
and the Higgs background included all Higgs decays other than the signal ones.

The final event samples were determined using a simultaneous selection on the first and second neural nets as specified in %%@
Table~\ref{cutstab}. The selection was performed separately for each channel and decay mode by finding an optimal point at which the signal %%@
cross section uncertainty was minimized. A detailed account of event selections and variable distributions for all decay modes in both %%@
channels is given in~\cite{yb,yb2,yb3}.

\begin{center}
\begin{table}
\begin{tabular}{| c | c | c | c | c |}
\hline
Decay Mode & \multicolumn{2}{|c|}{Neutrino} & \multicolumn{2}{|c|}{Hadronic} \\	
 & NN1 $>$ & NN2 $>$ & NN1 $>$ & NN2 $>$ \\ \hline
$c\bar{c}$ & 0.33 & 0.38 & 0.36 & 0.29 \\
$b\bar{b}$ & 0.81 & 0.92 & 0.64 & 0.88 \\
$gg$ & 0.71 & 0.87 & 0.10 & 0.27 \\ \hline
\end{tabular}
\caption{Neural net selections in the neutrino and hadronic channels.}
\label{cutstab}
\end{table}
\end{center}

\section{Branching Ratio Calculation}\label{res}

The branching ratio of the Higgs boson decay were calculated using events that passed the final neural network
selection. The calculation was done by normalising the signal cross section to the inclusive Higgs cross section, $\sigma_{ZH}$ = %%@
209.0$\pm$9.8 fb,
as determined in an independent recoil mass analysis performed for the SiD Letter of Intent~\cite{LoI}. The branching ratio is then given by

\begin{equation}
BR(H \rightarrow f\bar{f}) = \frac{\sigma_{H \rightarrow f\bar{f}}}{\sigma_{ZH} \times BR(Z \rightarrow q\bar{q},\nu\bar{\nu})}
\end{equation}

where $\sigma_{H \rightarrow ff}$ is the signal cross section and $BR(Z \rightarrow q\bar{q},\nu\bar{\nu})$ is the decay branching ratio of %%@
the $Z$ boson into two jets or neutrinos depending on the channel. $BR(Z \rightarrow q\bar{q},\nu\bar{\nu})$ values were taken from the %%@
generator tables.
The cross section was calculated as follows:

\begin{equation}
\sigma_{H \rightarrow f\bar{f}} = \frac{N}{\varepsilon L}
\end{equation}

where $N$ was the number of signal events after all selections, $\varepsilon$ was the total signal efficiency and $L$ was the integrated %%@
luminosity. The cross section uncertainty took into account statistical fluctuations in the number of both signal and background events. %%@
Calculation of $BR(H \rightarrow gg)$ follows exactly the same procedure.

Summary of the results obtained for both neutrino and hadronic channels is given in Table~\ref{finalvals}. It shows the number of events after %%@
all selections, signal efficiency, cross section and the branching ratios with their uncertainties for each channel separately and in %%@
combination. The cross sections and branching ratios, determined for combinations of neutrino and hadronic channels, and their relative %%@
uncertainties were calculated assuming that the two channels were statistically independent. The uncertainty of the cross section was %%@
considered to be of purely statistical nature while 
the uncertainty of the branching ratio also took into account the total Higgs-strahlung cross section uncertainty, which is correlated for %%@
both channels and is the largest source of systematic uncertainty in the analysis. Other systematic 
uncertainties originating from $BR(Z \rightarrow q\bar{q},\nu\bar{\nu})$ and luminosity uncertainties are of the order 
of 0.1\% and were neglected. The efficiency was derived using larger signal samples and was assumed not to contribute to the uncertainty. %%@
Systematic uncertainties originating from the detector properties, reconstruction algorithms and from the polarization measurement uncertainty %%@
were not addressed in this study though they are expected to be small.

\begin{center}
\begin{table*}[t]
\begin{tabular}{| c | c | c | c | c |}

\hline	
\multicolumn{2}{|c|}{~}
& ~Neutrino~ & ~Hadronic~ & ~Combined~ \\
\hline

~$H \rightarrow c\bar{c}$
~ & Signal events & 178 & 407 & ~ \\
~ & SM background events & 140 & 673 & ~ \\
~ & Higgs background events & 109 & 213 & ~ \\
~ & Signal efficiency & $27.9$ & $22.2$ & ~ \\
~ & Signal $\sigma_{H \rightarrow c\bar{c}}$ & $6.8 \pm 0.8$\,fb & $6.9 \pm 0.6$\,fb & $6.86 \pm 0.48$\,fb \\
~ & Relative uncertainty on $\sigma_{H \rightarrow c\bar{c}}$ & 11.6\% & 8.8\% & 7.0\% \\
~ &  Higgs BR & 3.3$\pm$0.4\% & 3.3$\pm$0.3\% & 3.3$\pm$0.3\% \\
~ & Relative uncertainty on Higgs BR & 12.5\% & 10.0\% & 8.4\% \\
\hline

~$H \rightarrow b\bar{b}$
~ & Signal events & 2833 & 8122 & ~ \\
~ & SM background events & 220 & 4700 & ~ \\
~ & Higgs background events & 55 & 423 & ~ \\
~ & Signal efficiency & $24.5$ & $26.2$ & ~ \\
~ & Signal $\sigma_{H \rightarrow b\bar{b}}$ & $142.7 \pm 2.3$\,fb & $142.5 \pm 1.9$\,fb & $142.57 \pm 1.61$\,fb \\
~ & Relative uncertainty on $\sigma_{H \rightarrow b\bar{b}}$ & 1.9\% & 1.4\% & 1.1\% \\
~ &  Higgs BR & 68.3$\pm$3.4\% & 68.2$\pm$3.3\% & 68.2$\pm$5.3\% \\
~ & Relative uncertainty on Higgs BR & 5.0\% & 4.9\% & 4.8\% \\
\hline
	
~$H \rightarrow gg$
~ & Signal events & 32 & 524 & ~ \\
~ & SM background events & 0 & 3621 & ~ \\
~ & Higgs background events & 4 & 1431 & ~ \\
~ & Signal efficiency & $3.3$ & $17.7$ & ~ \\
~ & Signal $\sigma_{H \rightarrow gg}$ & $15.1 \pm 1.9$\,fb & $15.6 \pm 2.6$\,fb & $15.41 \pm 1.74$\,fb \\
~ & Relative uncertainty on $\sigma_{H \rightarrow gg}$ & 18.7\% & 14.2\% & 11.3\% \\
~ &  Higgs BR & 7.2$\pm$1.4\% & 7.5$\pm$1.1\% & 7.4$\pm$0.9\% \\
~ & Relative uncertainty on Higgs BR & 19.3\% & 15.0\% & 12.2\% \\
\hline

\end{tabular}
\caption{Results for the $H\rightarrow c\bar{c}$, $H\rightarrow b\bar{b}$ and $H\rightarrow gg$ decay modes.}
\label{finalvals}
\end{table*}
\end{center}
The branching ratios were measured as 3.3\%, 68.2\% and 7.4\% for $c\bar{c}$, $b\bar{b}$ and $gg$ respectively. It was found that deviations %%@
of the determined branching ratios from the values used in the MC generator were  within the statistical uncertainties in all cases implying %%@
that the analysis did not bias the measurement. The relative uncertainties on the branching ratios were determined to be equal to 8.4\%, 4.8\% %%@
and 12.2\% for $c\bar{c}$, $b\bar{b}$ and $gg$ respectively. For the $c\bar{c}$ and gg measurements, the relative uncertainty of the signal %%@
cross section dominated the uncertainty of the branching ratio. However, for the main decay channel, $b\bar{b}$, the branching ratio %%@
uncertainty was dominated by the uncertainty on the inclusive Higgs-strahlung cross section used for normalization.

The obtained results were found largely consistent with results from earlier studies in \cite{snowmass}, \cite{tesla} and \cite{ZH}, which all %%@
used simplified approaches either to the simulation, or reconstruction, or both. The deterioration of sensitivity due to the added realism was %%@
recovered with a new signal discrimination technique which utilizes correlations of two neural networks.

\section{Summary}\label{summary}

The sensitivity to the decay branching ratios of a neutral 120 GeV SM Higgs boson to charm quarks, bottom quarks and gluons has been studied %%@
at the ILC centre-of-mass energy of $\sqrt{s}$ = 250 GeV and integrated luminosity of 250 fb$^{-1}$. The analysis is based on full simulation %%@
and realistic event reconstruction in the SiD detector and also fully accounts for the standard model background processes. The relative %%@
uncertainties obtained are comparable to values obtained in some of the previous studies. Good performance of flavour tagging and the use of %%@
neural networks in event selection were critical in obtaining these results. The uncertainties on the branching ratios were found to be equal %%@
to 8.4\%, 4.8\% and 12.2\% for $c\bar{c}$, $b\bar{b}$ and $gg$, respectively. The uncertainty in the $b\bar{b}$ branching ratio is dominated %%@
by the uncertainty on the inclusive Higgs-strahlung cross section.

\begin{acknowledgments}
We would like to thank colleagues from the SiD software and benchmarking groups, in particular Jan Strube, Tim Barklow, 
Norman Graf, and John Jaros for assistance with processing and useful discussions.
\end{acknowledgments} 

\newpage

\bibliography{hcbg}

%Merlin.mbs v4.21 2009-07-09.
\hyphenation{Post-Script Sprin-ger}
\begin{thebibliography}{10}%
\makeatletter
\providecommand \@ifxundefined [1]{%
 \ifx #1\undefined \expandafter \@firstoftwo
 \else \expandafter \@secondoftwo
\fi
}%
\providecommand \@ifnum [1]{%
 \ifnum #1\expandafter \@firstoftwo
 \else \expandafter \@secondoftwo
\fi
}%
\providecommand \enquote [1]{``#1''}%
\providecommand \bibnamefont  [1]{#1}%
\providecommand \bibfnamefont [1]{#1}%
\providecommand \citenamefont [1]{#1}%
\providecommand\href[0]{\@sanitize\@href}%
\providecommand\@href[1]{\endgroup\@@startlink{#1}\endgroup\@@href}%
\providecommand\@@href[1]{#1\@@endlink}%
\providecommand \@sanitize [0]{\begingroup\catcode`\&12\catcode`\#12\relax}%
\@ifxundefined \pdfoutput {\@firstoftwo}{%
 \@ifnum{\z@=\pdfoutput}{\@firstoftwo}{\@secondoftwo}%
}{%
 \providecommand\@@startlink[1]{\leavevmode\special{html:<a href="#1">}}%
 \providecommand\@@endlink[0]{\special{html:</a>}}%
}{%
 \providecommand\@@startlink[1]{%
  \leavevmode
  \pdfstartlink
   attr{/Border[0 0 1 ]/H/I/C[0 1 1]}%
   user{/Subtype/Link/A<</Type/Action/S/URI/URI(#1)>>}%
  \relax
 }%
 \providecommand\@@endlink[0]{\pdfendlink}%
}%
\providecommand \url  [0]{\begingroup\@sanitize \@url }%
\providecommand \@url [1]{\endgroup\@href {#1}{\urlprefix}}%
\providecommand \urlprefix [0]{URL }%
\providecommand \Eprint[0]{\href }%
\@ifxundefined \urlstyle {%
  \providecommand \doi [1]{doi:\discretionary{}{}{}#1}%
}{%
  \providecommand \doi [0]{doi:\discretionary{}{}{}\begingroup
  \urlstyle{rm}\Url }%
}%
\providecommand \doibase [0]{http://dx.doi.org/}%
\providecommand \Doi[1]{\href{\doibase#1}}%
\providecommand \bibAnnote [3]{%
  \BibitemShut{#1}%
  \begin{quotation}\noindent
    \textsc{Key:}\ #2\\\textsc{Annotation:}\ #3%
  \end{quotation}%
}%
\providecommand \bibAnnoteFile [2]{%
  \IfFileExists{#2}{\bibAnnote {#1} {#2} {\input{#2}}}{}%
}%
\providecommand \typeout [0]{\immediate \write \m@ne }%
\providecommand \selectlanguage [0]{\@gobble}%
\providecommand \bibinfo [0]{\@secondoftwo}%
\providecommand \bibfield [0]{\@secondoftwo}%
\providecommand \translation [1]{[#1]}%
\providecommand \BibitemOpen[0]{}%
\providecommand \bibitemStop [0]{}%
\providecommand \bibitemNoStop [0]{.\EOS\space}%
\providecommand \EOS [0]{\spacefactor3000\relax}%
\providecommand \BibitemShut [1]{\csname bibitem#1\endcsname}%
%</preamble>
\bibitem{Higgs}%
  \BibitemOpen
  \bibfield{author}{%
  \bibinfo {author} {\bibfnamefont{P.}~\bibnamefont{Higgs}},\ }%
  \bibfield{journal}{%
  \bibinfo {journal} {Phys. Rev. Lett.}\ }%
  \textbf{\bibinfo {volume} {13}},\ \bibinfo {pages} {508} (\bibinfo {year}
  {1964})%
  \bibAnnoteFile{NoStop}{Higgs}%
\bibitem{lhcilc}%
  \BibitemOpen
  \bibfield{author}{%
  \bibinfo {author} {\bibfnamefont{G.}~\bibnamefont{Weiglein}}\ and\ \bibinfo
  {author} {\bibnamefont{{\it et al.}~(LHC/ILC Study~Group)}},\ }%
  \bibfield{journal}{%
  \bibinfo {journal} {Phys. Rept.}\ }%
  \textbf{\bibinfo {volume} {426}},\ \bibinfo {pages} {47} (\bibinfo {year}
  {2006})%
  \bibAnnoteFile{NoStop}{lhcilc}%
\bibitem{yukawa}%
  \BibitemOpen
  \bibfield{author}{%
  \bibinfo {author} {\bibfnamefont{A.}~\bibnamefont{Djouadi}}, \bibinfo
  {author} {\bibfnamefont{M.}~\bibnamefont{Spira}},\ and\ \bibinfo {author}
  {\bibfnamefont{P.}~\bibnamefont{Zerwas}},\ }%
  \bibfield{journal}{%
  \bibinfo {journal} {Phys. Lett.}\ }%
  \textbf{\bibinfo {volume} {B 264}} (\bibinfo {year} {1991})%
  \bibAnnoteFile{NoStop}{yukawa}%
\bibitem{mssm2}%
  \BibitemOpen
  \bibfield{author}{%
  \bibinfo {author} {\bibfnamefont{M.}~\bibnamefont{Battaglia}},\ }%
  \enquote{\bibinfo {title} {Measuring higgs branching ratios and telling the
  sm from a mssm higgs boson at the e+e- linear collider},}\ \bibinfo {note}
  {ArXiv:{h}ep-ph/9910271}%
  \bibAnnoteFile{NoStop}{mssm2}%
\bibitem{mssm1}%
  \BibitemOpen
  \bibfield{author}{%
  \bibinfo {author} {\bibfnamefont{M.}~\bibnamefont{Carena}}, \bibinfo {author}
  {\bibfnamefont{H.}~\bibnamefont{Haber}}, \bibinfo {author}
  {\bibfnamefont{H.}~\bibnamefont{Logan}},\ and\ \bibinfo {author}
  {\bibfnamefont{S.}~\bibnamefont{Mrenna}},\ }%
  \enquote{\bibinfo {title} {Distinguishing a mssm higgs boson from the sm
  higgs boson at a linear collider},}\ \bibinfo {note}
  {ArXiv:{h}ep-ph/0106116}%
  \bibAnnoteFile{NoStop}{mssm1}%
\bibitem{ZH}%
  \BibitemOpen
  \bibfield{author}{%
  \bibinfo {author} {\bibfnamefont{T.}~\bibnamefont{Kuhl}}\ and\ \bibinfo
  {author} {\bibfnamefont{K.}~\bibnamefont{Desch}},\ }%
  \enquote{\bibinfo {title} {Simulation of the measurement of the hadronic
  branching ratios for a light higgs boson at the ilc},}\ \bibinfo {note}
  {{L}C-PHSM-2007-001}%
  \bibAnnoteFile{NoStop}{ZH}%
\bibitem{LoI}%
  \BibitemOpen
  \emph{\bibinfo {title} {Si{D} {L}etter of {I}ntent}},\ edited by\ \bibinfo
  {editor} {\bibfnamefont{H.}~\bibnamefont{Aihara}}, \bibinfo {editor}
  {\bibfnamefont{P.}~\bibnamefont{Burrows}},\ and\ \bibinfo {editor}
  {\bibfnamefont{M.}~\bibnamefont{Oreglia}}\ (\bibinfo {year} {2009})\ \bibinfo
  {note} {arXiv:0911.0006v1}%
  \bibAnnoteFile{NoStop}{LoI}%
\bibitem{zhggstr}%
  \BibitemOpen
  \bibfield{author}{%
  \bibinfo {author} {\bibfnamefont{B.}~\bibnamefont{Lee}}, \bibinfo {author}
  {\bibfnamefont{C.}~\bibnamefont{Quigg}},\ and\ \bibinfo {author}
  {\bibfnamefont{H.}~\bibnamefont{Thacker}},\ }%
  \bibfield{journal}{%
  \bibinfo {journal} {Phys. Rev.}\ }%
  \textbf{\bibinfo {volume} {D 16}},\ \bibinfo {pages} {1519} (\bibinfo {year}
  {1977})%
  \bibAnnoteFile{NoStop}{zhggstr}%
\bibitem{zhggstr2}%
  \BibitemOpen
  \bibfield{author}{%
  \bibinfo {author} {\bibfnamefont{J.}~\bibnamefont{Ellis}}, \bibinfo {author}
  {\bibfnamefont{M.}~\bibnamefont{Gaillard}},\ and\ \bibinfo {author}
  {\bibfnamefont{D.}~\bibnamefont{Nanopoulos}},\ }%
  \bibfield{journal}{%
  \bibinfo {journal} {Nucl. Phys.}\ }%
  \textbf{\bibinfo {volume} {B 106}},\ \bibinfo {pages} {292} (\bibinfo {year}
  {1976})%
  \bibAnnoteFile{NoStop}{zhggstr2}%
\bibitem{fuse}%
  \BibitemOpen
  \bibfield{author}{%
  \bibinfo {author} {\bibfnamefont{R.}~\bibnamefont{Cahn}}\ and\ \bibinfo
  {author} {\bibfnamefont{S.}~\bibnamefont{Dawson}},\ }%
  \bibfield{journal}{%
  \bibinfo {journal} {Phys. Lett.}\ }%
  \textbf{\bibinfo {volume} {B 136}},\ \bibinfo {pages} {196} (\bibinfo {year}
  {1984})%
  \bibAnnoteFile{NoStop}{fuse}%
\bibitem{whiz}%
  \BibitemOpen
  \bibfield{author}{%
  \bibinfo {author} {\bibfnamefont{W.}~\bibnamefont{Kilian}}, \bibinfo {author}
  {\bibfnamefont{T.}~\bibnamefont{Ohl}},\ and\ \bibinfo {author}
  {\bibfnamefont{J.}~\bibnamefont{Reuter}},\ }%
  \enquote{\bibinfo {title} {Whizard: Simulating multiple particle processes at
  lhc and ilc},}\ \bibinfo {note} {ArXiv:0708.4233}%
  \bibAnnoteFile{NoStop}{whiz}%
\bibitem{pyth}%
  \BibitemOpen
  \bibfield{author}{%
  \bibinfo {author} {\bibfnamefont{T.}~\bibnamefont{Sjostrand}}, \bibinfo
  {author} {\bibfnamefont{S.}~\bibnamefont{Mrenna}},\ and\ \bibinfo {author}
  {\bibfnamefont{P.}~\bibnamefont{Skands}},\ }%
  \bibfield{journal}{%
  \bibinfo {journal} {JHEP}\ }%
  \textbf{\bibinfo {volume} {026}} (\bibinfo {year} {2006})%
  \bibAnnoteFile{NoStop}{pyth}%
\bibitem{geant}%
  \BibitemOpen
  \bibfield{author}{%
  \bibinfo {author} {\bibfnamefont{S.}~\bibnamefont{Agostinelli}}\ and\
  \bibinfo {author} {\bibnamefont{{\it et al}}},\ }%
  \bibfield{journal}{%
  \bibinfo {journal} {Nucl. Instrum. Methods Phys. Res.}\ }%
  \textbf{\bibinfo {volume} {A 506}},\ \bibinfo {pages} {250} (\bibinfo {year}
  {2003})%
  \bibAnnoteFile{NoStop}{geant}%
\bibitem{geant2}%
  \BibitemOpen
  \bibfield{author}{%
  \bibinfo {author} {\bibfnamefont{J.}~\bibnamefont{Allison}}\ and\ \bibinfo
  {author} {\bibnamefont{{\it et al}}},\ }%
  \bibfield{journal}{%
  \bibinfo {journal} {IEEE Trans. Nucl. Sci.}\ }%
  \textbf{\bibinfo {volume} {53}},\ \bibinfo {pages} {270} (\bibinfo {year}
  {2006})%
  \bibAnnoteFile{NoStop}{geant2}%
\bibitem{slic}%
  \BibitemOpen
  \enquote{\bibinfo {title} {Slic},}\ \bibinfo {note}
  {Http://www.lcsim.org/software/slic/doxygen/html/}%
  \bibAnnoteFile{NoStop}{slic}%
\bibitem{lcio}%
  \BibitemOpen
  \bibfield{author}{%
  \bibinfo {author} {\bibfnamefont{F.}~\bibnamefont{Gaede}}, \bibinfo {author}
  {\bibfnamefont{T.}~\bibnamefont{Behnke}}, \bibinfo {author}
  {\bibfnamefont{N.}~\bibnamefont{Graf}},\ and\ \bibinfo {author}
  {\bibfnamefont{T.}~\bibnamefont{Johnson}},\ }%
  \enquote{\bibinfo {title} {Lcio - a persistency framework for linear collider
  simulation studies},}\ \bibinfo {note} {{L}C-TOOL-2003-053}%
  \bibAnnoteFile{NoStop}{lcio}%
\bibitem{durham}%
  \BibitemOpen
  \bibfield{author}{%
  \bibinfo {author} {\bibfnamefont{S.}~\bibnamefont{Catani}}, \bibinfo {author}
  {\bibfnamefont{Y.}~\bibnamefont{Dokshitzer}}, \bibinfo {author}
  {\bibfnamefont{M.}~\bibnamefont{Olsson}}, \bibinfo {author}
  {\bibfnamefont{G.}~\bibnamefont{Turnock}},\ and\ \bibinfo {author}
  {\bibfnamefont{B.}~\bibnamefont{Webber}},\ }%
  \bibfield{journal}{%
  \bibinfo {journal} {Phys. Lett.}\ }%
  \textbf{\bibinfo {volume} {B 268}},\ \bibinfo {pages} {432} (\bibinfo {year}
  {1991})%
  \bibAnnoteFile{NoStop}{durham}%
\bibitem{kinfit}%
  \BibitemOpen
  \bibfield{author}{%
  \bibinfo {author} {\bibfnamefont{B.}~\bibnamefont{List}}\ and\ \bibinfo
  {author} {\bibfnamefont{J.}~\bibnamefont{List}},\ }%
  \enquote{\bibinfo {title} {Marlinkinfit: An object-oriented kinematic fitting
  package},}\ \bibinfo {note} {{L}C-TOOL-2009-001}%
  \bibAnnoteFile{NoStop}{kinfit}%
\bibitem{lcfi}%
  \BibitemOpen
  \bibfield{author}{%
  \bibinfo {author} {\bibfnamefont{A.}~\bibnamefont{Bailey}}\ and\ \bibinfo
  {author} {\bibnamefont{{\it et al.}~(LCFI~Collaboration)}},\ }%
  \bibfield{journal}{%
  \bibinfo {journal} {Nucl. Instrum. Methods Phys. Res.}\ }%
  \textbf{\bibinfo {volume} {A 610}},\ \bibinfo {pages} {573} (\bibinfo {year}
  {2009})%
  \bibAnnoteFile{NoStop}{lcfi}%
\bibitem{thrs}%
  \BibitemOpen
  \bibfield{author}{%
  \bibinfo {author} {\bibfnamefont{S.}~\bibnamefont{Brandt}},\ }%
  \emph{\bibinfo {title} {Jet Analysis in Electron-Positron Annihilation
  Experiments}}\ (\bibinfo {publisher} {Springer Berlin},\ \bibinfo {year}
  {1985})%
  \bibAnnoteFile{NoStop}{thrs}%
\bibitem{fann}%
  \BibitemOpen
  \enquote{\bibinfo {title} {Fast {A}rtificial {N}eural {N}etwork {L}ibrary
  (fann)},}\ \bibinfo {note} {Http://leenissen.dk/fann/html/files/fann-h.html}%
  \bibAnnoteFile{NoStop}{fann}%
\bibitem{yb}%
  \BibitemOpen
  \bibfield{author}{%
  \bibinfo {author} {\bibfnamefont{Y.}~\bibnamefont{Banda}}, \bibinfo {author}
  {\bibfnamefont{T.}~\bibnamefont{Lastovicka}},\ and\ \bibinfo {author}
  {\bibfnamefont{A.}~\bibnamefont{Nomerotski}},\ }%
  \enquote{\bibinfo {title} {Measurement of the higgs boson decay branching
  ratio to charm quarks at the ilc},}\ \bibinfo {note} {ArXiv:0909.1052v3}%
  \bibAnnoteFile{NoStop}{yb}%
\bibitem{yb2}%
  \BibitemOpen
  \bibfield{author}{%
  \bibinfo {author} {\bibfnamefont{Y.}~\bibnamefont{Banda}}, \bibinfo {author}
  {\bibfnamefont{T.}~\bibnamefont{Lastovicka}},\ and\ \bibinfo {author}
  {\bibfnamefont{A.}~\bibnamefont{Nomerotski}},\ }%
  \enquote{\bibinfo {title} {Precision measurement of higgs decay branching
  ratios to bottom quarks and gluons at the ilc},}\ \bibinfo {note}
  {ArXiv:1003.1333v1}%
  \bibAnnoteFile{NoStop}{yb2}%
\bibitem{yb3}%
  \BibitemOpen
  \bibfield{author}{%
  \bibinfo {author} {\bibfnamefont{Y.}~\bibnamefont{Banda}},\ }%
  \emph{\bibinfo {title} {Determination of the Higgs Boson Branching Ratios at
  the International Linear Collider}},\ Ph.D. thesis,\ \bibinfo {school}
  {University of Oxford} (\bibinfo {year} {2010})%
  \bibAnnoteFile{NoStop}{yb3}%
\bibitem{snowmass}%
  \BibitemOpen
  \bibfield{author}{%
  \bibinfo {author} {\bibfnamefont{C.~T.}\ \bibnamefont{Potter}}, \bibinfo
  {author} {\bibfnamefont{J.~E.}\ \bibnamefont{Brau}},\ and\ \bibinfo {author}
  {\bibfnamefont{M.}~\bibnamefont{Iwasaki}},\ }%
  in\ \emph{\bibinfo {booktitle} {Proc. of the APS/DPF/DPB Summmer Study on the
  Future of Particle Physics (Snowmass 2001)}},\ \bibinfo {series and number}
  {eConf C010630},\ \bibinfo {editor} {edited by\ \bibinfo {editor}
  {\bibfnamefont{N.}~\bibnamefont{Graf}}}\ (\bibinfo {year} {2001})\ p.\
  \bibinfo {pages} {118}%
  \bibAnnoteFile{NoStop}{snowmass}%
\bibitem{tesla}%
  \BibitemOpen
  \bibfield{author}{%
  \bibinfo {author} {\bibfnamefont{R.}~\bibnamefont{Brinkmann}}\ and\ \bibinfo
  {author} {\bibnamefont{{\it et al}}},\ }%
  \enquote{\bibinfo {title} {Tesla technical design report},}\ \bibinfo {note}
  {{D}ESY-2001-011}%
  \bibAnnoteFile{NoStop}{tesla}%
\end{thebibliography}%

\end{document}